\documentstyle[12pt]{article}

\newcommand{\ep}{\varepsilon}
\newcommand{\vp}{\varphi}

\newcommand{\bea}{\begin{eqnarray}}
\newcommand{\eea}{\end{eqnarray}}
\newcommand{\ps}{\varphi}
\begin{document}

\setcounter{page}{0} \thispagestyle{empty}

\vskip 2cm

\begin{center}
{\Large {\bf 
DGLAP and BFKL evolution equations in the $N=4$
supersymmetric gauge theory. 
}} \\[0pt]
\vspace{1.5cm} {\large \ A.V. Kotikov
}
\\[0pt]
\vspace{0.5cm} {\em Bogoliubov Laboratory of Theoretical Physics \\[0pt]
Joint Institute for Nuclear Research\\[0pt]
141980 Dubna, Russia }\\[0pt]

\vspace{0.5cm} and \\[0pt]
\vspace{0.5cm} {\large \ L.N. Lipatov}
\\[0pt]
\vspace{0.5cm} {\em Theoretical Physics Department\\[0pt]
Petersburg Nuclear Physics Institute\\[0pt]
Orlova Roscha, Gatchina\\[0pt]
188300, St. Petersburg
}\\[0pt]
\end{center}

\vspace{1.5cm}\noindent

\begin{center}
{\bf Abstract}
\end{center}

We discuss DGLAP and BFKL evolution equations in the $N=4$
supersymmetric gauge theory in the leading and next-to-leading
approximations. Eigenvalues of the BFKL kernel in this model turn
out to be analytic functions of the conformal spin. It allows us
to find the residues of the anomalous dimensions of
the twist-2 operators in the points $j=1,0,-1,....$ from the BFKL
equation in an agreement with their direct calculation from the DGLAP
equation. The holomorphic separability of the BFKL kernel and the
integrability of the DGLAP dynamics in this model are also discussed.  

\newpage


\pagestyle{plain}

\section{Introduction}

\indent

The Balitsky-Fadin-Kuraev-Lipatov (BFKL) equation \cite{BFKL, BL} is used
now together with the Dokshitzer-Gribov-Lipatov-Altarelli-Parisi (DGLAP)
equation \cite{DGLAP} for a theoretical description of structure functions
of the deep-inelastic $ep$ scattering at small values of the Bjorken
variable $x$. In this kinematical region the structure functions are
measured by the H1 and ZEUS collaborations \cite{H1}. The radiative 
corrections to the splitting kernels for the DGLAP equation are well known
\cite{corAP}. Although the BFKL equation in the leading logarithmic
approximation (LLA) was constructed many years ago, the calculation of the
next-to-leading corrections to its kernel was started only in 1989 
\cite{LF89}
and completed comparatively recently \cite{FL,CaCi,KoLi}. 

In supersymmetric gauge theories the structure of BFKL and DGLAP equations
is simplified significantly. 
In the case of 
the extended $N=4$ SUSY model the next-to-leading order (NLO) corrections
to BFKL
equation were calculated in ref. \cite{KoLi} for arbitrary values of the
conformal spin $n$ in a framework of the dimensional regularization 
(DREG) scheme.
In the section 3 below the results are presented in 
the scheme of the dimensional reduction
(DRED) which  does not violate  supersymmetry.    
The analyticity of the eigenvalue  of the BFKL kernel over the
conformal
spin
$n$ gives a possibility
to  relate DGLAP and BFKL  
equations in this
model, as we show below. Further, the eigenvalue  in both schemes
(DREG and DRED) can be written as a sum of the eigenvalues of holomorphic
and anti-holomorphic operators (see Section 3). 
\\

Let us to introduce 
the unintegrated parton 
distributions (UnPD) $\vp_a(x,k^2_{\bot})$ (hereafter $a=q,g,\ps$ for the 
quark
and gluon densities respectively) and the (integrated) parton 
distributions (PD) $f_a(x,Q^2)$, where
\bea
f_a(x,Q^2) ~=~ \int_{k^2_{\bot} < Q^2} dk^2_{\bot} \vp_a(x,k^2_{\bot}).
\label{in1}
\eea

The DGLAP equation relates the (integrated) parton distributions
having different values of $Q^2$. It has the form:
\bea
\frac{d}{d\ln{Q^2}} f_a(x,Q^2) &=& - \tilde W_a f_a(x,Q^2) +
\int_{x}^1 \frac{dy}{y} \sum_{b} \tilde W_{b\to a}(x/y) f_b(y,Q^2), 
\nonumber \\
\tilde W_a &=& \sum_b \int^1_0 dx \,\,x\,\, \tilde W_{a\to b}(x),
\label{in2}
\eea
where the last term in the r.h.s. is the Mellin convolution of the
inclusive
transition probabilities
$\tilde W_{b\to a}(x)$ and PD $f_b(x,Q^2)$.
Usually the first and second terms in the right-hand side of the
equation are unified by modifying 
$\tilde W_{b\to a}$ in the form of the splitting kernel
$W_{b\to a}$:
\bea
\frac{d}{d\ln{Q^2}} f_a(x,Q^2) &=& 
\int_{x}^1 \frac{dy}{y} \sum_{b} W_{b\to a}(x/y) f_b(y,Q^2)
\label{in2.1}
\eea

It is known, that DGLAP equation (\ref{in2.1}) is simplified essentially
after
the Mellin transformation to the $t$-channel angular momentum $j$
representation:
\bea
\frac{d}{d\ln{Q^2}} f_a(j,Q^2) &=& 
\sum_{b} \gamma_{ab}(j) f_b(j,Q^2),
\label{in2.2}
\eea
where 
\bea
f_a(j,Q^2) = \int^1_0 dx \,\, x^{j-1}f_a(x,Q^2)
\label{in2.21}
\eea
 are the Mellin moments
of parton distributions. The Mellin moment of the splitting kernel
\bea
\gamma_{ab}(j) = \int^1_0 dx \,\, x^{j-1}W_{b\to a}(x)
\label{in2.22}
\eea
coincides with
the anomalous dimension matrix for twist-2 operators 
\footnote{ As in Ref. \cite{FL,KoLi}, the anomalous dimensions differ 
from ones used usually in DIS by a factor $(-2)$, i.e. 
$\gamma_{ab}(j)=(-1/2)\gamma_{ab}^{DIS}(j)$}. These operators are
constructed as bilinear combinations of 
derivations of the fields describing the partons $a$
 (see Eq. (\ref{in4}) below).

\vskip 0.3cm

The BFKL equation relates the unintegrated parton distributions
having various values of the Bjorken variable $x$ and has the form:
\bea
\frac{d}{d\ln{(1/x)}} \vp_g(x,k^2_{\bot}) ~=~  
2 \omega(-k^2_{\bot})\,\vp_g(x,k^2_{\bot}) +
\int d^2k'_{\bot} K(k_{\bot},k'_{\bot}) 
\vp_g(x,k^2_{\bot}),
\label{in3}
\eea
where $ \omega(-k^2_{\bot})$ is the gluon Regge trajectory.\\

Let us introduce the local twist-two operators:
\bea
O_{\mu_1,...,\mu_j}^g &=& \hat S G_{\rho \mu_1}\,D_{\mu_2}\,D_{\mu_3}...
\,D_{\mu_{j-1}}G_{\rho \mu_j}, \nonumber \\
\tilde O_{\mu_1,...,\mu_j}^g &=& \hat S G_{\rho \mu_1}\,D_{\mu_2}\,D_{\mu_3}...
\,D_{\mu_{j-1}}\tilde G_{\rho \mu_j}, \nonumber \\
O_{\mu_1,...,\mu_j}^q &=& \hat S \overline{\Psi}
\gamma_{\mu_1}\,D_{\mu_2}... \,D_{\mu_j} \Psi, \nonumber \\
\tilde O_{\mu_1,...,\mu_j}^q &=& \hat S \overline{\Psi}
\gamma_{5}\,\gamma_{\mu_1}\,D_{\mu_2}... \,D_{\mu_j} \Psi, \nonumber \\
O_{\mu_1,...,\mu_j}^{\ps} &=& \hat S \Phi
D_{\mu_1}\,D_{\mu_2}... \,D_{\mu_j} \Phi, 
\label{in4}
\eea
where the last operator is constructed from derivations of the scalar
field
$\Phi $ appearing in supersymmetric models. The symbol $\hat S$ implies
a symmetrization of the tensor in the Lorenz indices $\mu_1,...,\mu_j$ and
a subsequent  
subtraction of its traces.

The matrix elements of $O_{\mu_1,...,\mu_j}^a$ and 
$\tilde O_{\mu_1,...,\mu_j}^a$ are
related to the moments of parton distributions
in a hadron $h$ in the following way
\bea
\int_{0}^1 dx x^{j-1} f_a(x,Q^2) &=& 
<h| \tilde n^{\mu_1}... \, \tilde n^{\mu_j}\, O_{\mu_1,...,\mu_j}^a |h>,
~~~~~~~~ a=(q,g,\ps),
\nonumber  \\
\int_{0}^1 dx x^{j-1} \Delta f_a(x,Q^2) &=& 
<h| \tilde n^{\mu_1}...\,\tilde n^{\mu_j}\, \tilde O_{\mu_1,...,\mu_j}^a |h>,
~~~~~~~~ a=(q,g),
\label{in5}
\eea
where the vector $\tilde n^{\mu}$ is light-like: $\tilde n^{2}=0$.
Note, that in the deep-inelastic $ep$ scattering 
$\tilde n^{\mu}=q+ x p$.\\

The 
conformal spin $n$ and the quantity $1+\omega $, expressed in terms of the
eigenvalue $\omega$ 
of the BFKL kernel,
coincide respectively with the total numbers of  transverse and 
longitudinal indices of the Lorentz tensor with the rank
$j=1+\omega +n$. 
Namely, we can introduce the following projectors of this  tensor

\bea
n^{\mu_1}...n^{\mu_{j-|n|}}\, O_{\mu_1,...,\mu_{j-|n|}\,
\sigma_1,...,\sigma_{|n|}}^a 
\, l^{\sigma_1}_{\bot}...l^{\sigma_{|n|}}_{\bot},
\label{in5}
\eea
where the complex transverse vector $l_{\bot}$ is given below 
$$
l^{\sigma}_{\bot}~=~ \frac{1}{\sqrt{2}} \left(\delta^{\sigma}_1 + 
i\delta^{\sigma}_2 \right),~~~l^2_{\bot}=0 \,.
$$

Anomalous dimension matrices $\gamma_{ab}(j)$ and $\tilde \gamma_{ab}(j)$
for
the twist-2 operators 
$O_{\mu_1,...,\mu_j}^a$ and $\tilde O_{\mu_1,...,\mu_j}^a$ 
do not depend on the different projections of the tensors due to the
Lorentz invariance.

In the gluon case the usual light-cone projections entering in the DGLAP
equation 
(see Eqs.(\ref{in5})) are 
\bea
\tilde n^{\mu_1}...\,\tilde n^{\mu_j}\, <P| O_{\mu_1,...,\mu_j}^g |P> ~=~
\int_{0}^1 dx x^{j-1} 
f_g(x,Q^2).
\label{in5.5}
\eea

The mixed
projections 
\bea
\tilde n^{\mu_1}...\, \tilde n^{\mu_{1+\omega}}\, 
l^{\mu_{1+\omega}}_{\bot}...l^{\mu_j}_{\bot}\,
 <P| O^{\mu_1,...,\mu_j}_g |P> ~\sim ~
\int_{0}^1 dx x^{\omega} 
\int d^2k_{\bot} {\left(\frac{k_{\bot}}{|k_{\bot}|}\right)}^n 
\vp_g(x,k^2_{\bot}),
\label{in6}
\eea
can be expressed in terms
of the solution of the BFKL equation. Generally the corresponding operators 
have  higher
twists.\\

Thus, it looks possible to obtain
some additional information about the
parton $x$-distributions
satisfying the DGLAP equation from the analogous
$k_{\bot}$-distributions satisfied the BFKL equation. Moreover, in the 
extended $N=4$ SUSY the $\beta$-function 
equals zero and therefore the $4$-dimensional conformal invariance could
allow to  relate the
Regge and Bjorken asymptotics of scattering amplitudes.\\


Our presentation is organized as follows. In Section 2 we discuss the
relation between DGLAP and BFKL equations in the leading logarithmic 
approximation.
In Section 3 we review shortly the
results of Ref. \cite{KoLi} and rewrite them 
in the framework of
the DRED
scheme. 
Section 4 contains the information about the 
anomalous dimensions calculated independently with the use
of the renormalization group. 
A summary is given in Conclusion. 


\section{Anomalous dimensions of twist-2 operators and their singularities}

\indent

In the leading logarithmic approximation (LLA) for the BFKL equation the 
contribution of fermions is not essential and therefore in this approximation
the integral kernel is the same for all supersymmetric gauge theories.
In the impact parameter representation due to the conformal invariance the
solution of 
the homogeneous BFKL equation has the form (see \cite{conf})

\bea 
E_{\nu,n}(\overrightarrow{\rho _{10}},
\overrightarrow{\rho _{20}}) \equiv
<\,\phi (\overrightarrow{\rho _{1}}) 
\, O_{m,\tilde m}(\overrightarrow{\rho _{0}})\,
\phi (\overrightarrow{\rho _{2}})\,>\,
~=~\left(\frac{\rho _{12}}{\rho _{10} 
\rho _{20}}\right)^m
\left(\frac{\rho _{12}^*}{\rho _{10}^* \rho 
_{20}^*}\right)^{\widetilde{m}},
\label{2.1}
\eea
where
\bea
m=1/2+i\nu +n/2,\,\,\widetilde{m}=1/2+i\nu -n/2.
\nonumber 
\eea
are the conformal weights. 
Here we introduced the complex variables
$\rho_{k}=x_k+iy_k$ and denoted $\rho_{kl}=\rho_{k}-\rho_{l}$

For the principal series of the unitary representations the quantities
$\nu$ and $n$ are
correspondingly
real and integer numbers. The projection $n$ of the
conformal spin $|n|$ can be positive or negative, 
but the eigenvalue of the BFKL equation in LLA
\bea
\omega =\omega^0(n,\nu)&=& \frac{g^2N_c}{8\pi^2}
\biggl(4\Psi \Bigl(1\Bigr)-\Psi \Bigl(\frac{1}{2}+i\nu +\frac{n}{2}\Bigr)-
\Psi \Bigl(\frac{1}{2}-i\nu -\frac{n}{2}\Bigr) \nonumber \\
&-&\Psi \Bigl(\frac{1}{2}+i\nu -\frac{n}{2}\Bigr)-
\Psi \Bigl(\frac{1}{2}-i\nu +\frac{n}{2}\Bigr) \biggr)
\label{2.2}
\eea
depends only on $|n|$. We shall imply below that $n$ is positive, i.e.
\bea
n~=~|n|\,.
\label{2.3}
\eea
Note, that eq. (\ref{2.2}) has the property of the holomorphic
separability and corresponds to the pair hamiltonian of the integrable
Heisenberg spin model (see \cite{integr}-\cite{Lipatov}). 

The solution of the inhomogeneous BFKL equation in the LLA approximation
can be written as the four-point function of a two-dimensional field
theory
\bea 
<\,\phi (\overrightarrow{\rho _{1}}) \, \phi (\overrightarrow{\rho _{2}})\,
\phi (\overrightarrow{\rho_{1'}}) \, \phi (\overrightarrow{\rho_{2'}})\,>\,
=~
\sum_n \int _{-\infty}^{\infty} d \nu C(\nu,n)
\int d^2 \rho _0
\,\frac{ E_{\nu,n}(\overrightarrow{\rho _{10}},
\overrightarrow{\rho _{20}})
E^*_{\nu,n}(\overrightarrow{\rho _{1'0}},
\overrightarrow{\rho _{2'0}})}{\omega -\omega^0 (n,\nu)},
\label{2.4}
\eea
where $C(\nu,n)$ is expressed through the inhomogeneous term of the
equation with the use 
of the completeness condition for $E_{\nu , n}$ (see \cite{conf}).

For $\overrightarrow{\rho_{1'}} \to \overrightarrow{\rho_{2'}}$
we obtain from the integration region
$\overrightarrow{\rho_{0}} \to \overrightarrow{\rho_{1'}}$:
\bea 
<\,\phi (\overrightarrow{\rho _{1}}) \, \phi (\overrightarrow{\rho _{2}})\,
\phi (\overrightarrow{\rho_{1'}}) \, \phi (\overrightarrow{\rho_{2'}})\,>\,
&\sim &
\sum_n \int _{-\infty}^{\infty} d \nu C(\nu,n)
\,\frac{ E_{\nu,n}(\overrightarrow{\rho _{11'}},
\overrightarrow{\rho _{21'}})}{\omega -\omega^0 (n,\nu)}
\rho^m_{1'2'}\rho^{* \tilde m}_{1'2'} \nonumber \\
&\sim &
\sum_n  C(\nu_{\omega},n)
\,\frac{ E_{\nu_{\omega},n}(\overrightarrow{\rho _{11'}},
\overrightarrow{\rho _{21'}})}{\omega' (n,\nu_{\omega})}
{|\rho_{1'2'}|}^{1+2i\nu_{\omega}}
{\left(\frac{\rho_{1'2'}}{\rho^{*}_{1'2'}}\right)}^{n/2}
\,
\label{2.5}
\eea
where $\nu_{\omega}$ is a solution of the equation
\bea
\omega =\omega^0(n,\nu)
\label{2.6}
\eea
with $Im \nu_{\omega}<0$.

The above asymptotics has a simple interpretation in terms of the Wilson
operator-product expansion
\bea 
\lim_{\rho_{1'} \to \rho_{2'}}
\phi (\overrightarrow{\rho _{1'}}) \, \phi (\overrightarrow{\rho_{2'}})\,
~= ~
\sum_n  
\,\frac{C(\nu_{\omega},n) }{\omega' (n,\nu_{\omega})}
{|\rho_{1'2'}|}^{2\Gamma_{\omega}}
{\left(\frac{\rho_{1'2'}}{\rho^{*}_{1'2'}}\right)}^{n/2}
O_{\nu_{\omega},n}(\overrightarrow{\rho _{1'}})
\,,
\label{2.7}
\eea
where 
\bea
\Gamma_{\omega} =\frac{1}{2} + i\nu_{\omega}
\label{2.8}
\eea
is the
transverse dimension of the  operator 
$O_{\nu_{\omega},n}(\overrightarrow{\rho _{1'}})$, calculated in units 
of squared mass. This operator is the following projection
\bea
O_{\nu_{\omega},n}(\overrightarrow{\rho _{1'}}) ~=~
\tilde{n}^{\mu_1}...\tilde{n}^{\mu_{1+\omega}}\, 
l^{\sigma_1}_{\bot}...l^{\sigma_n}_{\bot}\,
 O_{\mu_1,...,\mu_{1+\omega}\sigma_1,...,\sigma_n} 
\label{2.9}
\eea
of the gauge-invariant tensor with $1+\omega +n$ indices.

The anomalous dimension $\gamma(\omega)$ obtained from the BFKL equation
in LLA has the poles
\bea
\Gamma_{\omega} =1+\frac{|n|}{2} - i\gamma (j),~~
\gamma (j)|_{\omega \to 0}= \frac{g^2N_c}{4\pi^2 \omega}.
\label{2.10}
\eea

The canonical contribution $1+|n|/2$ to the transverse dimension
$\Gamma_{\omega}$
corresponds to the local operator
\bea
G_{\rho\mu_1} D_{\mu_2}^{||}...D_{\mu_{\omega}}^{||}
D_{\sigma_1}^{\bot}...D_{\sigma_n}^{\bot} G_{\rho\mu_{1+\omega}},
\label{2.11}
\eea
because in the light-cone gauge $A_{\mu} \tilde{n}_{\mu}=0$ the tensor
\bea
G_{\rho\mu_1}G_{\rho\mu_{1+\omega}}\,
\tilde{n}^{\mu_1}\,\tilde{n}^{\mu_{2}}
~\sim ~
 \partial _{\mu_1}A^{\bot}_{\rho} \partial _{\mu_2}A^{\bot}_{\rho}\,
\tilde{n}^{\mu_1}\,\tilde{n}^{\mu_{2}}
\label{2.12}
\eea
has the transverse dimension equal to 1. 

The above local operator
$O_{\nu _{\omega},n}$ for $|n|>0$ should have the twist higher than 2
because its anomalous
dimension is singular at $\omega \to 0$. 
Indeed, such singularities are impossible for the 
twist-2 operator, because for $n>0$ and small $\omega$ the total number of
its indices is
integer and physical.
The only way to obtain some information about the twist-2
operators
is  to continue analytically the anomalous dimension $\gamma(\omega)$
to negative integer points
\bea
|n| ~\to ~ -r-1,
\label{2.13}
\eea
where $r$ is a positive integer.
Then in the limit $\omega \rightarrow 0$ the total number of indices
$j=1+\omega+|n|$ tends to $-r$
and therefore the pole of $\gamma (\omega)$ in $\omega$ can be interpreted 
as the pole $1/(j+r)$ for a non-physical value of $j$ for the twist-2
operator.

In LLA one can  obtain after the analytic continuation of 
$\omega ^0 (|n|, \nu )$
\bea
\gamma (j)|_{j \to -r}~=~ \frac{g^2N_c}{4\pi^2} \frac{1}{j+r}
\label{2.14}
\eea
We remind, that for the BFKL equation in LLA the fermions are not important,
but generally they give non-vanishing contributions to the residues of the
poles of
$\gamma (j)$ in the DGLAP equation even in LLA.
Therefore the above result for $\gamma (j)$ can be  valid only for a
definite generalization of QCD. It is known \cite{KoLi}, that only for the
extended
N=4 supersymmetric Yang-Mills theory the anomalous dimension, calculated
in the
next-to-leading approximation for the BFKL equation, 
can be analytically continued to $|n|=-r-1$. Therefore it is natural to 
expect, that the above result for $\gamma (j)$ in LLA is valid for the N=4
case.

Indeed, using the conservation of the stress tensor in this theory to fix 
the substruction constant in the expansion over the poles at $j=-r$, we
obtain :
\bea
\gamma (j)^{LLA}~=~ \frac{g^2N_c}{4\pi^2} \Bigl(\Psi(1)-\Psi(j-1)\Bigr) 
\label{2.15}
\eea
in an agreement with the direct calculation of $\gamma (j)$ in this theory
(see \cite{N=0,Dubna} and Section 4).

In the next-to-leading order (NLO)
approximation it is needed to modify the above 
procedure of the derivation of $\gamma (j)$ from the BFKL equation, taking
into account the possibility of the 
appearance of double-logarithmic terms leading to  triple poles at
$j=-k$.
We shall return to this problem in our future publication.


\section{NLO corrections to BFKL kernel in $N=4$ SUSY }

\indent

To begin with, we review shortly the results of Ref. \cite{KoLi}, where the 
NLO corrections to the BFKL integral kernel at $t=0$ were calculated in
the case of QCD and
 supersymmetric gauge theories.
We discuss only the case of the $N=4$
supersymmetric gauge theory and write 
the formulae important for our analysis.\\

\subsection{The set of eigenvalues}

The set of eigenvalues for eigenfunctions 
of the homogeneous BFKL
equation for the $N=4$ supersymmetric theory 
\bea
\omega =
4\,\overline{a}\,
\biggl[ \chi (n,\gamma)
+\left(\frac{1}{3} \chi (n,\gamma) + \delta (n,\gamma ) \right)
\,\overline{a} \,
\biggr] \, \label{KL1}
\eea
has been found in  \cite{KoLi}
in the following form 
\begin{eqnarray}
\chi (n,\gamma ) &=&2\Psi (1)-\Psi \Bigl(\gamma +\frac{n}{2}\Bigr)-\Psi
\Bigl%
(1-\gamma +\frac{n}{2}\Bigr)  \label{7} \\
&&  \nonumber \\
\delta (n,\gamma )&=&-\Biggl[ 2\Phi (n,\gamma )+2\Phi
(n,1-\gamma ) 
+2\zeta (2)
\chi (n,\gamma )  \nonumber \\
&-& 6\zeta (3)- 
 \Psi ^{\prime \prime }\Bigl(\gamma +\frac{n}{2}\Bigr)- \Psi ^{\prime
\prime } \Bigl(1-\gamma +\frac{n}{2}\biggr) \Biggr],  \label{K3}
\end{eqnarray}
where 
$\Psi (z)$, $\Psi
^{\prime }(z)$ and $\Psi ^{\prime \prime }(z)$ are respectively the Euler
$\Psi $
-function and its  derivatives; $\overline{a}=g^2N_c/(16\pi^2)$
is coupling constant in DREG scheme. The function $\Phi (n,\gamma)$ 
is given below\footnote{Notice that the representation of $\Phi (n,\gamma)$
contains a misprint in \cite{KoLi}: the factor
$(-1)^{k+1}$ in the last sum of (\ref{9}) was substituted
by $(-1)^{k}$ }
\begin{eqnarray}
\Phi (n,\gamma ) &=&-\int_{0}^{1}dx~
\frac{x^{\gamma -1 +n/2}}{1+x}\Biggr[ \frac{1%
}{2}\biggl( \Psi ^{\prime }\Bigl(\frac{n+1}{2}\Bigr)-\zeta (2)\biggr) +{\rm
Li}_{2}(-x)+{\rm Li}_{2}(x)  \nonumber \\
&+&\ln (x)\biggl(\Psi (n+1)-\Psi (1)+\ln (1+x)+\sum_{k=1}^{\infty }\frac{%
(-x)^{k}}{k+n}\biggr)  \nonumber \\
&+&\sum_{k=1}^{\infty }\frac{x^{k}}{(k+n)^{2}}(1-(-1)^{k})\Biggr]  \nonumber
\\
&&\hspace*{-1cm}\hspace{-1.3cm}=~\sum_{k=0}^{\infty }\frac{(-1)^{k+1}}{%
k+\gamma +n/2}\Biggl[ \Psi ^{\prime }(k+1)-\Psi ^{\prime }(k+n+1)+(-1)^{k+1}%
\Bigl(\beta ^{\prime }(k+n+1)+\beta ^{\prime }(k+1)\Bigr)\biggr)  \nonumber
\\
&+&\frac{1}{k+\gamma +n/2}\biggl( \Psi (k+n+1)-\Psi (k+1)%
\biggr) \Biggr],  \label{9}
\end{eqnarray}
and
\[
\beta ^{\prime }(z)=\frac{1}{4}\Biggl[ \Psi ^{\prime }\Bigl(\frac{z+1}{2}
\Bigr)-\Psi ^{\prime }\Bigl(\frac{z}{2}\Bigr)\Biggr],~~~
\beta ^{\prime \prime }(z)=\frac{1}{8}
\Biggl[ \Psi ^{\prime \prime}\Bigl(\frac{z+1}{2}
\Bigr)-\Psi ^{\prime \prime}\Bigl(\frac{z}{2}\Bigr)\Biggr],~~~
\]

Note, that the term 
\bea
\frac{1}{3} \chi (n,\gamma )
\label{9.1}
\eea
appears as a result of the use of the 
DREG scheme 
(see (29) in \cite{KoLi}). 
It is well known, that DREG  violates SUSY. The proper
procedure, which is very close to DREG  and satisfies the 
supersymmetry requirements, 
is the DRED scheme. The results in the framework of the
DRED scheme
\footnote{In the calculations given in 
\cite{KoLi} the result obtained in the DRED scheme with $6$ scalars and
pseudoscalars is equal
to that for DREG  with $6+2\hat\ep$ scalars and pseudoscalars,
where $\hat\ep=(4-D)/2$ and $D$ is the space-time dimension.}
can be found from (\ref{KL1})
%
by the redefinition
of the coupling constant 
\bea
{\overline{a}}\to \hat a={\overline{a}}+
\frac{1}{3}
{\overline{a}}^2,
\label{9.1a}
\eea
which eliminates the above term (\ref{9.1}) in (\ref{KL1}).
For the new coupling constant $a$ the above expression for $\omega$ can be
written in
the following form
\bea
\omega =
4\, \hat a \,
\biggl[ \chi (n,\gamma )+\delta(n,\gamma )
\, \hat a \,
\biggr]
\label{9.1b}
\,
\eea
\vskip 0.5cm

\subsection{Generalized holomorphical separability}

Following the method of Refs. \cite{N=0,Dubna} and using the results of the
previous section we can 
present the above NLO corrections to the BFKL equation in the form 
having a (generalized) holomorphical separability.
We 
can split the function $\Phi (n,\gamma)$ in two parts

\[
\Phi (n,\gamma )=\Phi _1(n,\gamma )+\Phi _2(n,\gamma )\,,
\]
where
\[
\Phi _1(n,\gamma )=~\sum_{k=0}^\infty \frac{\left( \beta ^{\prime
}(k+n+1)-(-1)^k\Psi ^{\prime }(k+n+1)\right) }{k+\gamma +n/2}
\]
\[
+\sum_{k=0}^\infty \frac{(-1)^k\left( \Psi (n+k+1)-\Psi (1)\right) }{%
(k+\gamma +n/2)^2}\,,
\]
and

\[
\Phi _2(n,\gamma )=~\sum_{k=0}^\infty \frac{\left( \beta ^{\prime
}(k+1)+(-1)^k\Psi ^{\prime }(k+1)\right) }{k+\gamma +n/2}
\]
\[
-\sum_{k=0}^\infty \frac{(-1)^k\left( \Psi (k+1)-\Psi (1)\right) }{(k+\gamma
+n/2)^2} \equiv \Phi _2(\gamma +n/2 )\,\,.
\]
Here $\Phi _2(n,\gamma )$ depends only on $m=\gamma +n/2$ and therefore 
the corresponding contributions to $\omega$ have the property of the
holomorphic
separability.  Further, for $\Phi _1(n,\gamma )$ we obtain the simple
relation
\[
\Phi _1(n,\gamma )+\Phi _1(n,1-\gamma )~=~\beta ^{\prime }(\gamma
+n/2)\left[ \Psi (1)-\Psi (1-\gamma +n/2)\right]
\]
\[
+\beta ^{\prime }(1-\gamma +n/2)\left[ \Psi (1)-\Psi (\gamma +n/2)\right]
\,,
\]
which can be verified by calculating the residues at $\gamma =-r-n/2$ and
$\gamma
=1+r+n/2
$ and the asymptotic behavior at $\gamma \rightarrow \infty $. Therefore
one has
\[
\Phi (n,\gamma )+\Phi (n,1-\gamma )=\chi (n,\gamma )\,\left( \beta ^{\prime
}(\gamma +\frac n2)+\beta ^{\prime }(1-\gamma +\frac n2)\right) +\Phi
_2(\gamma +\frac n2)-
\]
\[
\beta ^{\prime }(\gamma +\frac n2)\left[ \Psi (1)-\Psi (\gamma +\frac
n2)\right] +\Phi _2(1-\gamma +\frac n2)-\beta ^{\prime }(1-\gamma +\frac
n2)\left[ \Psi (1)-\Psi (1-\gamma +\frac n2)\right],
\]
where $\chi (n,\gamma )$ is given by Eq.(\ref{7}).

Thus, we can rewrite the NLO corrections $\delta(n,\gamma)$ in 
the generalized holomorphic separable form (providing that $\omega _0$ is 
substituted by $\omega$, which is valid with our accuracy):

\begin{eqnarray}
\delta (n,\gamma )&=& \phi\Bigl(\gamma +\frac{n}{2}\Bigr) + 
\phi\Bigl(1-\gamma +\frac{n}{2}\Bigr) - 
\frac{\omega_0}{2\hat a} \biggl(\rho\Bigl(\gamma +\frac{n}{2}\Bigr) + 
\rho\Bigl(1-\gamma +\frac{n}{2}\Bigr)\biggr) 
\label{7.0} \\
\omega_0 &=& 4\hat a \biggl(2\Psi (1)-\Psi \Bigl(\gamma +\frac{n}{2}\Bigr)-\Psi
\Bigl
(1-\gamma +\frac{n}{2}\Bigr)\biggr)  \label{7.1}
\end{eqnarray}
and
\begin{eqnarray}
\rho (\gamma ) &=&\beta^{(1)} (\gamma )+\frac{1}{2} \zeta(2)  \label{7.2} \\
&&  \nonumber \\
\phi (\gamma )&=& 3\zeta(3) + \Psi^{('')}(\gamma) - 2\Phi_2(\gamma) 
+2\beta^{(1)} (\gamma )\Bigl(2\Psi (1)-\Psi (\gamma)\Bigr)\,.
 \label{7.3}
\end{eqnarray}


\subsection{The asymptotics of ``cross-sections'' at $s\to \infty$}

\indent

As an example we consider
the cross-sections for the inclusive production of two pairs of 
particles with mass $m$ 
in the polarised $\gamma \gamma$ collision (see \cite{BL,KoLi}):
\[
\sigma (s)~=~
\alpha^2 _{em} \hat a^2
\frac{1}{m^{2}}\frac{32}{81}\biggl(%
\sigma _{0}(s)+\Bigl(\cos ^{2}\vartheta -\frac{1}{2}
\Bigr)\sigma _{2}(s)\biggr),
\]
where $\alpha _{em}$ is the electromagnetic coupling constant,  
the coefficient $\sigma _{0}(s)$ is proportional to the cross-section for
the scattering of unpolarized photons and $\sigma _{2}(s)$ describes
the spin correlation depending on the relative azimutal angle
$\vartheta $ between the polarization vectors of colliding photons.

The asymptotic behavior of the cross-sections $\sigma _{k}(s)~~(k=0,2)$ at
$s\to \infty $ leads in the $t$-channel
to unmoving 
singularities $f_{\omega }(t) \sim (\omega -\omega _{k})^{1/2}$, where
\bea
\omega _{k}&=& 4\hat a
\Biggl[
\chi (k,\frac{1}{2})+ \hat  a
\delta(k,\frac{1}{2}) \Biggr]
\equiv 4\hat a
\chi (k,\frac{1}{2})\Biggl[1- \hat a
c(k,\frac{1}{2})\Biggr]  ~~~\left(c(n,\gamma) = 
-\frac{\delta (n,\gamma)}{\chi (n,\gamma)} \right),
\nonumber \\
& & \nonumber \\
\sigma _{0}(s) &=&\frac{9\pi ^{5/2}}{32\sqrt{7\zeta (3)}}\frac{s^{\omega
_{0}}}{{\Bigl(\ln(s/s_0) 
\Bigr)}^{1/2}}\cdot \biggl(1+O(a)\biggr), \\
\sigma _{2}(s) &=&\frac{\pi ^{5/2}}{9\cdot 32\sqrt{7\zeta (3)-8}}\frac{%
s^{\omega _{2}}}{{\Bigl(\ln(s/s_0) 
\Bigr)}^{1/2}}\cdot \biggl(1+O(a)\biggr).
\end{eqnarray}
Here the symbol $O(\hat a)$ denotes unknown next-to-leading corrections
to the impact factors.

Using our results (26), (27), we obtain in the $N=4$ case the
following values for $\chi (k,%
\frac{1}{2})$ and $c(k,\frac{1}{2})$ $(k=0,2)$:
\begin{eqnarray}
\chi (0,\frac{1}{2}) &=&4\ln 2,~~~
\chi (2,\frac{1}{2})~=~4(\ln2-1)\,,  \label{10} \\
&&  \nonumber \\
c(0,\frac{1}{2}) &=&
2\zeta(2)
+\frac{1}{2\ln 2}\Biggl[ 11\zeta (3)-32{\rm Ls}_{3}\Big(\frac{\pi }{2}\Big)
-\frac{165}{16}
\pi \zeta (2)
\Biggr] ~=~9.5812
\,,
\label{11} \\
&&  \nonumber \\
c(2,\frac{1}{2}) &=&
2\zeta (2)
+\frac{1}{2(\ln 2-1)}  
\Biggl[ 11\zeta (3)+32{\rm Ls}_{3}\Big(\frac{\pi }{2}\Big)+
14 \pi \zeta (2)-32\ln 2\Biggr]
\nonumber \\ 
&=& 6.0348
\,,
\label{12}
\end{eqnarray}
where (see \cite{Lewin,Devoto})
\[
{\rm Ls}_{3}(x)=-\int_{0}^{x}\ln ^{2}\left| 2\sin \Bigl(\frac{y}{2}\Bigr%
)\right| dy \,.
\]
Note, that the function 
${\rm Ls}_{3}(x)$ appears also in  calculations of some
massive diagrams (see, for example, the recent papers \cite{LS3} and
references therein).

The LO results (\ref{10}) coincide with ones obtained in ref. \cite{BL}. 
As it was shown in \cite{FL,KoLi}, in the framework of QCD 
the NLO correction $c^{QCD}(0,1/2)$ is
large and leads to a quite strong reduction of the value of the Pomeron
intercept (see recent analyses \cite{bfklp}-\cite{resum} of various effective
resummations of the large NLO terms). Contrary to $c^{QCD}(0,1/2)$, the
correction $c(0,1/2)$ is not large ($c^{QCD}(0,1/2)/c(0,1/2) \approx 2.7$),
which seems to support the results of Ref \cite{bfklp}, where a quite
large
value
of the non-conformal contribution to $c^{QCD}(0,1/2)$ was found in
the physical renormalization
schemes.
The values of $c^{QCD}(2,1/2)$ (see \cite{FL,KoLi}) and $c(2,1/2)$
are small and do not change significantly the small LO value
\cite{BL} of the angle-dependent contribution.

\subsection{Non-symmetric choice of the energy normalization}

Analogously to 
refs. \cite{FL,KoLi} one can calculate 
the eigenvalues of the kernel in the case
of a non-symmetric 
choice of the energy normalization parameter $s_0$ in eq.(15).
For the scale $s_0=q^2$, which is natural for the deep-inelastic
scattering process, 
we obtain

in DREG-scheme
\begin{eqnarray}
\omega = 4\, \overline{a}\,
\Biggl[ \chi(n,\gamma )+ \biggl(\frac{1}{3} \chi (n,\gamma )+
\delta(n,\gamma ) - 2 \chi(n,\gamma ) \chi'(n,\gamma )
\biggl)
\, \overline{a} \,
\Biggr]
\label{K4.1}
\end{eqnarray}

and in DRED-scheme
\begin{eqnarray}
\omega = 4\, \hat a \,
\Biggl[ \chi(n,\gamma )+ \biggl(
\delta (n,\gamma ) - 2 \chi(n,\gamma ) \chi'(n,\gamma )
\biggl)
\, \hat a \,
\Biggr],
\label{K4.1a}
\end{eqnarray}
where
$$
\chi'(n,\gamma )\equiv \frac{d}{d\gamma} \chi(n,\gamma )
= -\Psi ^{\prime }\Bigl(\gamma +\frac{n}{2}\Bigr)+\Psi ^{\prime}
\Bigl(1-\gamma +\frac{n}{2}\biggr)
$$

\subsection{The limit $\gamma \to 0$  for $n=0$}

By considering the limit $\gamma \to 0$ in the Eq.(\ref{K4.1a})
we have for $n=0$ (see also the analysis in \cite{KoLi})
\begin{eqnarray}
\chi(0,\gamma ) &=& \frac{1}{\gamma} + O(\gamma^2), ~~ \nonumber \\
\frac{1}{3} \chi (n,\gamma ) &+&
 \delta(0,\gamma ) - 2 \chi(0,\gamma ) \chi'(0,\gamma )
~=~ \frac{B^{DREG}}{\gamma}
+ C + O(\gamma^2), \nonumber \\
 \delta(0,\gamma ) &-& 2 \chi(0,\gamma ) \chi'(0,\gamma )
 ~=~ 
\frac{B^{DRED}}{\gamma} + C + O(\gamma^2), \label{n1}
\end{eqnarray}
where

\begin{eqnarray}
 B^{DREG} ~=~ \frac{1}{3} ,~~~
B^{DRED} ~=~ 0, ~~~\mbox{ and }~~~
C ~=~  2\zeta(3).
\label{n7}
\end{eqnarray}

Analogous to ref.\cite{FL,KoLi} with the use of eqs.(\ref{n1}) and
(\ref{n7})
one can obtain the expression for 
anomalous dimensions of twist-2 operators $\gamma$ at $\omega \to 0$
(i.e. near $j=1$)

in DREG-scheme
\begin{eqnarray}
\gamma = 4\, \overline{a}\, \Biggl[
\biggl( \frac{1}{\omega} + O(\omega)\biggr)
+ a \,
\biggl( \frac{ B^{DREG}}{\omega} + O(1) \biggr)
+ \overline{a}^2 \,
\biggl( \frac{C}{\omega^2} + O\left(\omega^{-1}\right) \biggr)
\Biggl]
\label{n5}
\end{eqnarray}

and in DRED-scheme
\begin{eqnarray}
\gamma =  4\, \hat a \, \Biggl[
\biggl( \frac{1}{\omega} + O(\omega)\biggr)
+ \hat a \,
\biggl( \frac{B^{DRED}}{\omega} + O(1) \biggr)
+ \hat a^2 \,
\biggl( \frac{C}{\omega^2} + O\left(\omega^{-1}\right) \biggr)
\Biggl].
\label{n6}
\end{eqnarray}

Thus,
in the framework of DRED scheme the vanishing contribution 
at $j \to 1$ is obtained for the  NLO contribution to the anomalous
dimension.\\

\subsection{Symmetry between $\gamma$ and $j-\gamma$}

Using  the scale $s_0=q^2$,
the expression for $\omega $ as a
function of the correctly defined anomalous dimension $\gamma $ at general
$n$ can be written in the following form
\bea
\omega =4\hat a \biggl( \chi\left(n,\gamma-\frac{n}{2}\right)
+ \hat a \biggl[\delta \left(n,\gamma-\frac{n}{2}\right)-
\chi\left(n,\gamma-\frac{n}{2}\right)\cdot 
\chi'\left(n,\gamma-\frac{n}{2}\right)
\biggr] \biggr),
\label{n7a}
\eea
where
\bea
\chi\left(n,\gamma-\frac{n}{2}\right) &=& 
2\,\Psi (1)-\Psi (\gamma )-\Psi (n+1-\gamma) \label{n7b}\\
\delta \left(n,\gamma-\frac{n}{2}\right) &=&
\Psi ^{\prime \prime }(\gamma )+
\Psi ^{\prime \prime}(n+1-\gamma )
-2\,\Phi \left(n,\gamma -\frac n2\right)
-2\,\Phi \left(n,1-\gamma +\frac n2\right) \nonumber \\
&+&
6\zeta (3) \,.
\nonumber
\eea

To calculate the anomalous dimension $\gamma $ we write $\omega$ in
the ''Lorentz invariant'' form

\bea
\omega =4\hat a\biggl( 2\,\Psi (1)-\Psi (\gamma )-\Psi (j
-\gamma )+\Delta (\,n,\,\gamma )\, \hat a\biggr) \,,\,\,j=n+1+\omega \,,
\label{n7d}
\eea
where
\bea
\Delta (\,n,\,\gamma )=\delta \left(n,\gamma-\frac{n}{2}\right)+
2\,\biggl[\Psi^{\prime }(\gamma )+\Psi^{\prime }(n+1-\gamma )\biggr]
 \chi\left(n,\gamma-\frac{n}{2}\right)\,.
\label{n7e}
\eea

Using the analysis of the subsection 3.2, Eq.(\ref{n7d}) can be presented  
as follows
\bea
\omega =4 \hat a\,\left( 2\,\Psi (1)-\Psi (\gamma )-\Psi (n+1+\omega -\gamma
)+\varepsilon \right) \,,\,\,
\label{n7f}
\eea
where $\varepsilon $ can be written in the ''holomorphically separable''
form 
\bea
\varepsilon &=&\frac{\omega }2\Bigl( p (\gamma )+p (1+n-\gamma
)\Bigr) + \hat a\Bigl( \phi (\gamma )+\phi (1+n-\gamma )\Bigr) ,
\nonumber \\
\,\omega &=& 4 \hat a
\Bigl( 2\,\Psi (1)-\Psi (\gamma )-\Psi (n+1-\gamma )\Bigr) +
O(\hat{a}^2)\,.
\label{n7g}
\eea

Here
\[
p (\gamma )=-\beta ^{\prime }(\gamma )+\Psi ^{\prime }(\gamma
)-\frac{1}{2}\zeta(2)~=~2\sum_{k=0}^\infty \frac 1{(\gamma +2k)^2}
-\frac{1}{2}\zeta(2)\,\,
\]
and $\phi (\gamma )$ is given by Eq.(\ref{7.3}).\\

Let us calculate $\Delta (\,n,\,\gamma )$ near its singularities. To
begin with, we consider  $\gamma \rightarrow 0$ for physical $n\geq 0$:
\[
\Delta (\,n,\,\gamma )\rightarrow \frac 4{\gamma ^2}\left(
\,\Psi (1)-\Psi (n+1)\right) \,+\frac 2\gamma \,c(n)\,,
\]

\[
c(n)=2\,\Psi ^{\prime }(n+1)-2\Psi ^{\prime }(1)-\beta ^{\prime }(n+1)-\beta
^{\prime }(1).
\]
Therefore
\[
\gamma =\frac{4\hat a}\omega \biggl( 1+\omega \left( \,\Psi (1)-
\Psi (n+1)\right)
\biggr) +\frac{(4\hat a)^2}{\omega ^2}\left( \Psi (1)-\Psi (n+1)+
\frac \omega 2c(n)\right) \,.
\]
At $n=0$ the correction $\sim \hat a$ is absent, 
but for other $n$, especially for $n\rightarrow -r-1$, we have the large 
correction to $\gamma$

\[
\Delta \gamma =4\hat a\,\left( \Psi (1)-\Psi (n+1)\right) \,,
\]
which has the pole at $j\rightarrow -r$ and leads to a contribution
changing even singularities of the Born term. It is related with the fact,
that for positive $n$ we calculate the anomalous dimensions of the
higher twist
operators (with the  anti-symmetrization
between $n$ transverse and $1+\omega $ longitudinal indices). \\

Because in the case $N=4$ SUSY the 
result is analytic in $|n|$, one can continue the anomalous dimensions 
to the negative values of $|n|$. It gives a possibility to find the singular
contributions of the anomalous dimensions of the twist-2 operators not 
only at $j=1$ but also at 
other integer points $j=0,\,-1,\,-2...$. In particular, as it was
discussed
above, in the Born 
approximation for the anomalous dimension of the supermultiplet of the
twist-2 operators we 
obtain $\gamma = 4\, \hat{a}\,
(\Psi (1)-\Psi (j-1))$ 
which coincides with the result of the direct calculations (see 
\cite{N=0,Dubna} and the discussion below). 
Thus, in the case $N=4$ the BFKL equation presumably contains 
the information sufficient for restoring the kernel of the
DGLAP equation.


\section{The anomalous dimensions matrix in the $N=4$ SUSY}

The DGLAP evolution equation for the 
 moments of the parton distributions for $N=4$ SUSY has the form
\bea
\frac{d}{d\ln{Q^2}} f_a(j,Q^2) &=&  \sum_{k} \gamma_{ab}(j) f_b(j,Q^2)
~~~~~~(a,b=q,g,\ps),
\label{3.0}\\
\frac{d}{d\ln{Q^2}} \Delta f_a(j,Q^2) &=&  \sum_{k} 
\tilde \gamma_{ab}(j) \Delta f_b(j,Q^2)
~~~~~~(a,b=q,g),
\label{3.01}
\eea
where the anomalous dimension matrices $\gamma_{ab}(j)$
an $\tilde \gamma_{ab}(j)$ can be written
as expansions over the  coupling constant $\hat a$
in the form:
\bea
\gamma_{ab}(j) ~=~ \hat a \cdot \gamma^{(0)}_{ab}(j) 
+ \hat a^2 \cdot \gamma^{(1)}_{ab}(j),~~~
\tilde \gamma_{ab}(j) ~=~ \hat a \cdot \tilde \gamma^{(0)}_{ab}(j) 
+ \hat a^2 \cdot \tilde \gamma^{(1)}_{ab}(j).
\label{3.1}
\eea

In the following subsections we will present the results of exact
calculations
for the leading order (LO) coefficients $\gamma^{(0)}_{ab}(j)$
and $\tilde \gamma^{(0)}_{ab}(j)$ and construct
the anomalous dimensions of the multiplicatively renormalizable operators
of
the twist-2.
In the NLO approximation the corresponding coefficients
$\gamma^{(1)}_{ab}(j)$
and $\tilde \gamma^{(1)}_{ab}(j)$
are yet unknown  and their calculation is in progress \cite{KoLiVe}.
However, the form of the LO anomalous dimensions of the multiplicatively 
renormalizable operators in $N=4$ SUSY is rather simple because they are 
expressed 
in terms of one function. Taking into account the universality of
this result, the knowledge of the anomalous dimensions
in the QCD case, the NLO corrections to the BFKL kernel
\cite{KoLi} and an experience in integrating some types of the Feynman
diagrams, one can write 
an ansatz for the NLO anomalous dimensions of the multiplicatively 
renormalizable operators in the $N=4$ SUSY.
This result will be checked by us later with direct calculations 
of the matrix elements $\gamma^{(1)}_{ab}(j)$ and 
$\tilde \gamma^{(1)}_{ab}(j)$.


\subsection{The results of exact calculations of the anomalous dimensions 
matrix in the $N=4$ SUSY}

The elements of the LO anomalous dimension matrix in the $N=4$ SUSY
have the following form (see \cite{Dubna}):\\

for usual tensor twist-2 operators 
\bea
\gamma^{(0)}_{gg}(j) &=& 4 \left( \Psi(1)-\Psi(j-1)-\frac{2}{j}+\frac{1}{j+1}
-\frac{1}{j+2} \right), \nonumber \\
\gamma^{(0)}_{qg}(j) &=& 8 \left(\frac{1}{j}-\frac{2}{j+1}+\frac{2}{j+2} 
\right),~~~~~~~~~~\,
\gamma^{(0)}_{\ps g}(j) ~=~ 12 \left( \frac{1}{j+1} -\frac{1}{j+2} \right),
\nonumber \\
\gamma^{(0)}_{gq}(j) &=& 2 \left(\frac{2}{j-1}-\frac{2}{j}+\frac{1}{j+1}
\right),~~~~~~~~~~\, 
\gamma^{(0)}_{q\ps}(j) ~=~ \frac{8}{j},
\nonumber \\
\gamma^{(0)}_{qq}(j) &=& 4 \left( \Psi(1)-\Psi(j)+\frac{1}{j}-
\frac{2}{j+1}\right),~~
\gamma^{(0)}_{\ps q}(j) ~=~ \frac{6}{j+1}, \nonumber \\
\gamma^{(0)}_{\ps \ps}(j) &=& 4 \left( \Psi(1)-\Psi(j+1)\right),
~~~~~~~~~~~~~~~\,
\gamma^{(0)}_{g\ps}(j) ~=~ 4 \left(\frac{1}{j-1}-\frac{1}{j} \right)
\label{3.2}
\eea

and for the pseudo-tensor operators:
\bea
\tilde \gamma^{(0)}_{gg}(j) &=& 4 \left( \Psi(1)-\Psi(j+1)-\frac{2}{j+1}
-\frac{2}{j+2} \right), \nonumber \\
\tilde \gamma^{a,(0)}_{qg}(j) &=& 8 \left(-\frac{1}{j}+\frac{2}{j+1}\right),
~~~~~
\tilde \gamma^{(0)}_{gq}(j) ~=~ 2\left( \frac{2}{j} -\frac{1}{j+1}\right),
\nonumber \\
\tilde \gamma^{(0)}_{qq}(j) &=& 4 \left( \Psi(1)-\Psi(j+1)+\frac{1}{j-1}-
\frac{1}{j}\right),
\label{3.3}
\eea

Note, that in $N=4$ SUSY there are also the twist-2 operators with the
fermion quantum numbers but the anomalous dimension of the
corresponding multiplicatively renormalizable operators is the same as that
for the bosonic components of the supermultiplet. 

\subsection{Anomalous dimensions and
twist-2 operators with a multiplicative renormalization}

Let us denote the distributions of partons with the spin $S$ by $f_S(x)$
and their corresponding momenta by $f_S(j)$. We introduce also the
inclusive probabilities $\Delta f_S(x)$ and their momenta $\Delta
f_S(j)$  which are the differences of the
distributions of the partons with the helicities $\pm S$.

It is possible to construct 5 independent twist-two operators with
the multiplicative renormalization. The corresponding parton distributions 
and their LO anomalous dimensions have the form (see \cite{Dubna}):

\bea
f_I(j) &=& f_g(j) + f_{q}(j) + f_{\ps}(j) \sim f^+_{q,g,\ps}(j), \nonumber \\ 
\gamma^{(0)}_{I}(j) &=& 4 \left( \Psi(1)-\Psi(j-1)\right) \equiv -4S_1(j-2) 
\equiv \gamma^{(0)}_{+}(j),
\nonumber \\
& &  \nonumber \\
f_{II}(j) &=& -2(j-1)f_g(j) + f_{q}(j) + \frac{2}{3}(j+1) f_{\ps}(j) 
\sim f^0_{q,g,\ps}(j), 
\nonumber \\ 
\gamma^{(0)}_{II}(j) &=& 4 \left( \Psi(1)-\Psi(j+1)\right)\equiv -4S_1(j)
\equiv \gamma^{(0)}_{0}(j), 
\nonumber \\
& &  \nonumber \\
f_{III}(j) &=& -\frac{j-1}{j+2}f_g(j) + f_{q}(j) - \frac{j+1}{j} f_{\ps}(j)
\sim f^-_{q,g,\ps}(j), 
\nonumber \\ 
\gamma^{(0)}_{III}(j) &=& 4 \left( \Psi(1)-\Psi(j+3)\right)\equiv -4S_1(j+2)
\equiv \gamma^{(0)}_{-}(j), 
\nonumber \\
& &  \nonumber \\
f_{IV}(j) &=& 2\Delta f_g(j) + \Delta f_{q}(j) \sim \Delta f^{+}_{q,g}(j), 
\nonumber \\ 
\gamma^{(0)}_{IV}(j) &=& 4 \left( \Psi(1)-\Psi(j)\right)\equiv -4S_1(j-1)
\equiv \tilde \gamma^{(0)}_{+}(j), 
\nonumber \\
& &  \nonumber \\
f_{V}(j) &=& -(j-1)\Delta f_1(j) + \frac{j+2}{2} \Delta f_{1/2}(j)
\sim \Delta f^{-}_{q,g}(j), 
\nonumber \\ 
\gamma^{(0)}_{IV}(j) &=& 4 \left( \Psi(1)-\Psi(j+2)\right)\equiv -4S_1(j+1)
\equiv \tilde \gamma^{(0)}_{-}(j),
\label{3.4}
\eea

Thus, we have one supermultiplet of operators with the same anomalous
dimension $\gamma ^{LO}(j)$, proportional to $\Psi (1) -\Psi (j-1)$. The 
momenta of the corresponding linear combinations of the parton
distributions can be obtained from the
above momenta by an appropriate shift of
their argument $j$ in accordance to  this universal anomalous 
dimension $\gamma ^{LO}(j)$. Moreover, the coefficients in these linear
combinations for $N=4$  SUSY can
be obtained from the conformal supersymmetry (cf. Ref
\cite{qp}) and should be the same for all orders of the
perturbation theory.\\

\subsection{The NLO anomalous dimensions and twist-two operators with
the multiplicative renormalization}


We have the following initial information to be able to predict the NLO
anomalous
dimensions of twist-two operators with
the multiplicative renormalization in N=4 SUSY:\\

{\bf 1.}~~ As it was shown in the previous subsections
LO anomalous dimensions are meromorphic functions. Moreover,
there is really only one basic anomalous dimension $\gamma^{LO}(j)$ and 
all other anomalous dimensions can be
obtained as $\gamma^{LO}(j \pm m)$, where $m$ is integer. 
It is useful to choose:

\bea
\gamma^{LO}(j) =
4\left(\Psi (1)-\Psi (j-1)\right) \equiv -4 S_1(j-2),
\label{4.1}
\eea
Then, $\gamma^{LO}(j)$ has a pole at $j \to 1$ and vanishes
at $j=2$.
It is natural to keep the above universality also for the NLO anomalous
dimensions $\gamma^{(1)}_{ab}(j)$ and $\tilde \gamma^{(1)}_{ab}(j)$. 
Adding some other
properties we will construct ansatz for the basic NLO anomalous
dimension $\gamma^{NLO}(j)$.\\

{\bf 2.}~~ There are well known results for NLO corrections to the QCD
anomalous
dimensions.\\

{\bf 3.}~~ 
In $\overline{MS}$-scheme (with the coupling constant $\overline a$) 
and also in
the $\overline{MS}$-like 
scheme with the coupling constant $ \hat a$ (i.e. in the scheme
based on DRED procedure), the terms 
$\sim \zeta(2)$ should be cancelled in the final result
for the forward Compton scattering (see \cite{CheKaTka}-\cite{Ko96}). 
Therefore the terms 
$\sim \zeta(2)$ should be cancelled at even $j$ in anomalous
dimensions for the structure functions $F_2$ and
$F_L$
(and for the unpolarized parton distributions) and at odd $j$ in anomalous
dimensions for structure functions $g_1$
and $F_3$
(and for the polarized  parton distributions).\\

{\bf 4.}~~ 
From the BFKL equation in the framework of DRED scheme
(see (\ref{9.1b}), (\ref{7}), (\ref{K3}) and
(\ref{9})) we know, that there is 
no mixing among the functions of different 'transcendentality level'
\footnote{Note that similar arguments have been used also in \cite{FleKoVe}
for finding results for some types of complicated massive
Feynman diagrams. Their evaluation was based
on the direct calculation of several terms in the series over the
inversed mass and on the knowledge of the basic structure of the series
obtained earlier in \cite{FleKoVe1,FleKoVe} by an exact calculation of few
special diagrams with the use of the differential equation method
\cite{DEM}.}
$i$, i.e. all special functions at the NLO correction contain the sums of
the
terms $\sim 1/n^i ~(i=3)$.
Indeed, if we denote the following  functions by the terms in the
corresponding sums
$$\Psi \sim 1/n,~~~ \Psi' \sim \beta' \sim \zeta(2) \sim 1/n^2,~~~
\Psi'' \sim \beta'' \sim \zeta(3) \sim 1/n^3,$$
then for the BFKL equation
the LO term and NLO one  have the 'levels' $i=1$
and $i=3$, respectively.

Because in N=4 SUSY there is a relation between BFKL and DGLAP equations,
the similar properties should be valid for anomalous dimensions
themselves,
i.e. the basic functions $\gamma^{LO}(j)$ and $\gamma^{NLO}(j)$
should be of the types $\sim 1/j^i$ with the levels $i=1$ and $i=3$,
respectively. 
The LO basic anomalous dimension is given by Eq. (\ref{4.1}).
Then, the
NLO basic anomalous dimension $\gamma^{NLO}(j)$
can be expressed through the functions:
$$S_i(j-2),~ K_i(j-2), S_{k,l}(j-2), K_{k,l}(j-2), 
\zeta(k) S_l(j-2),~ \zeta(k) K_i(j-2),~ \zeta(i)$$ (here $i=3$ and $j+l=i$),
where 
\bea
S_i(j)&=&\sum^{j}_{m=1}\frac{1}{m^i} \sim \Psi^{i-1}(j+1), \nonumber\\
S_{2,1}(j)&=& \sum^{j}_{m=1}\frac{1}{m^2}\,S_1(m),\nonumber \\
K_2(j)&=&\Bigl(1-(-1)^j \Bigr)\frac{1}{2}\,\zeta(2) +(-1)^j
\sum^{j}_{m=1}\frac{(-1)^{m+1}}{m^2} \sim \beta^{'}(j+1),\nonumber \\
K_3(j)&=&\Bigl(1-(-1)^j \Bigr)\frac{3}{4}\,\zeta(3) +(-1)^j
\sum^{j}_{m=1}\frac{(-1)^{m+1}}{m^3} \sim \beta^{''}(j+1),~~\nonumber \\
K_{2,1}(j)&=&\Bigl(1-(-1)^j \Bigr)\frac{5}{8}\,\zeta(3) +(-1)^j
\sum^{j}_{m=1}\frac{(-1)^{m+1}}{m^2}\,S_1(m)
\label{1.1}
\eea
Note 
that the terms $\sim \zeta(2)$ should be absent in accordance with item
{\bf 4}.\\

Moreover, the terms 
\bea
S_l(j-2)/(j \pm m)^k,~~~~ K_i(j-2)(j \pm m)^k ~~~~(j+l=i)
\label{1.1a}
\eea
 should be
absent, too. There are two reasons for this absence. 

First one, these
terms have additional poles at different values of $j$: $j =\mp m$.
But such additional poles should be absent, if we start 
with the BFKL equation and obtain $\gamma^{NLO}(j)$ by replacement 
$n \to -1-r$ because the BFKL equation does not dublicate the poles at
different
values
of $j$.  

The second reason comes from the consideration of linear combinations
(\ref{3.4}).
If, for example, in the polarized case terms (\ref{1.1a}) contribute,
then we will have the terms $\sim (j \pm m)^{-1}$ in one combination and
the terms $\sim (j \pm m \pm 2)^{-1}$ in another combination. However, 
from
the direct
calculation of QCD NLO anomalous dimensions in the polarized case 
(see \cite{MerNeer}) we
know that only the terms $\sim j^{-1}$ and $\sim (j + 1)^{-1}$ can 
contribute to these combinations. 

So, terms  (\ref{1.1a}) should be absent in the results for the 
universal NLO anomalous dimension in the
N=4 SUSY case.\\

{\bf 5.}~~
 Further, the NLO anomalous dimension $\gamma^{NLO}(j)$ is equal to a 
combination of
more complicated contributions (i.e. the contributions containing the
functions with
the maximal
value of $i$: $i=3$) for the QCD anomalous dimensions (with the SUSY 
relation for the QCD color factors $C_F=C_A=N_c$).

Note, that these most complicated contributions (with $i=3$) are 
the same for all
QCD anomalous dimensions
with above SUSY relation 
(with a possible exception for 
the NLO scalar-scalar anomalous dimension which is not known yet).\\

Thus, for $N=4$ SUSY 
the NLO universal anomalous dimension 
$\gamma^{NLO}(j)$ has the form
\bea
\gamma^{NLO}(j)~=~ 16 \, Q(j-2),
\label{1}
\eea
 where
\bea
Q(j) ~=~ K_{2,1}(j)+ \frac{1}{2} \biggl(
S_3(j)- K_3(j)\biggr)
+S_1(j) \biggl(S_2(j)-K_2(j) \biggr) 
\label{1a}
\eea
 
We would like to note that the sums $K_2(j)$, $K_3(j)$ and $K_{2.1}(j)$
(see Eq.(\ref{1.1}))
have been calculated directly
only at even values of $j$ (or at odd ones,
when $(-1)^j \to (-1)^{j+1}$). The analytical continuation to complex
values of $j$ 
can be done rather easily (see 
\cite{KaKo} and \cite{AnalCont}, respectively).\\

{\bf 6.}~~ 
We can add the term $\sim \zeta(3)$ to the r.h.s. of (\ref{1a}),  
but due to the condition\\ $\gamma^{NLO}(j=2)=0$ it cancels. \\

So, 
for N=4 SUSY the universal anomalous dimension $\gamma(j)$ has
the form 
\bea
\gamma(j)~=~ \hat a \gamma^{LO}(j) + \hat a^2 \gamma^{NLO}(j), ~~~~~~~
\label{2}
\eea
where $\gamma^{LO}(j)$ and $\gamma^{NLO}(j)$ are given by Eqs. (\ref{4.1})
and (\ref{1}), respectively. 
All other anomalous dimensions can be
obtained as $\gamma^{LO}(j \pm m)$ and 
$\gamma^{NLO}(j \pm m)$, where $m$ is an integer number. \\

Thus, the above arguments allow us to construct the NLO
corrections to anomalous dimensions 
in the N=4 SUSY, which were unknown earlier.
We plan, however, to check these results by direct calculations 
\cite{KoLiVe}.\\

\subsection{DGLAP evolution}

Using our knowledge of the anomalous dimensions we can construct the
solution of the DGLAP equation in the Mellin moment space in the framework
of 
N=4 SUSY.\\

{\bf A.~~Polarized case}

The polarized parton distributions are splitted in the two contributions:

\bea
\Delta f_{q,g}(j,Q^2) ~=~ \Delta f^{+}_{q,g}(j,Q^2) 
+ \Delta f^{-}_{q,g}(j,Q^2),
\label{AP1}
\eea
where

at LO
\bea
\Delta f^{\pm}_{q,g}(j,Q^2)~=~  \Delta f^{\pm,LO}_{q,g}(j,Q^2_0)
{\Biggl(\frac{Q^2}{Q^2_0}\biggr)}^{\tilde \gamma^{(0)}_{\pm}a}
~~~~~~~~~ \left(\tilde \gamma^{(0)}_{\pm}=4S_1(j\mp 1) \right)
\label{AP2}
\eea

at NLO
\bea
\Delta f^{\pm}_{q,g}(j,Q^2)~=~  \Delta f^{\pm,NLO}_{q,g}(j,Q^2_0)
{\Biggl(\frac{Q^2}{Q^2_0}\biggr)}
^{\left(\tilde \gamma^{(0)}_{\pm}a + \tilde \gamma^{(1)}_{\pm\pm}a^2\right)}
~~~~ \left(\tilde \gamma^{(1)}_{\pm\pm}=16 Q(j\mp 1) \right),
\label{AP2}
\eea
where
\bea
\Delta f^{\pm,NLO}_{q,g}(j,Q^2_0)&=& \Delta f^{\pm,NLO}_{q,g}(j,Q^2_0)
\Biggl(1- \frac{\tilde \gamma^{(1)}_{\pm\mp}\,\, \hat a}
{\tilde \gamma^{(0)}_{\pm}- \tilde\gamma^{(0)}_{\mp}}
\Biggr) \nonumber \\
&+& \frac{\tilde \gamma^{(1)}_{\mp\pm}\,\, \hat a}
{\tilde \gamma^{(0)}_{\mp}-\tilde \gamma^{(0)}_{\pm}}\,\,
\Delta f^{\mp,NLO}_{q,g}(j,Q^2_0)
\label{AP3}
\eea

Notice that
only 
the anomalous dimensions $\tilde \gamma^{(1)}_{\pm\pm}$ are important 
for $N=4$ SUSY  
at the order  $O(\hat a^2)$: they contribute to the $Q^2$-evolution of
parton distributions.

The anomalous dimensions $\tilde \gamma^{(1)}_{\pm\mp}$ 
give contributions at $O(\hat a^2)$ only to the normalization factors
$\Delta f^{\pm,NLO}_{q,g}(j,Q^2_0)$. Their contribution to the
$Q^2$-dependent part of $\Delta f^{\pm}_{q,g}(j,Q^2)$ 
starts at $O(\hat a^3)$
level in the following form:
\bea
\hat a^3 \,\,\frac{\tilde \gamma^{(1)}_{\pm\mp}\,\tilde\gamma^{(1)}_{\mp\pm}}
{\tilde \gamma^{(0)}_{\pm}-\tilde \gamma^{(0)}_{\mp}}
\label{AP4}
\eea

\vskip 0.5cm

{\bf B.~~Nonpolarized case} 

The polarized parton distributions are splitted in the three parts:

\bea
f_{q,g,\ps}(j,Q^2) ~=~ \sum_{i=+,-,0} f^{i}_{q,g,\ps}(j,Q^2), 
\label{AP1}
\eea
where\footnote{The arguments are $j-i2=\{j-2,j,j+2\}$ for respectively
$i=\{+,0,-\}$.}

at LO
\bea
f^{i}_{q,g,\ps}(j,Q^2)~=~  f^{i,LO}_{q,g,\ps}(j,Q^2_0)
{\Biggl(\frac{Q^2}{Q^2_0}\biggr)}^{\gamma^{(0)}_{i}a}
~~~~~~~~~ \left(\gamma^{(0)}_{i}=4S_1(j-i2) \right)
\label{AP2}
\eea

at NLO
\bea
f^{i}_{q,g,\ps}(j,Q^2)~=~  f^{i,NLO}_{q,g,\ps}(j,Q^2_0)
{\Biggl(\frac{Q^2}{Q^2_0}\biggr)}
^{\left(\gamma^{(0)}_{i}a + \gamma^{(1)}_{ii}a^2\right)}
~~~~ \left(\gamma^{(1)}_{ii}=16 Q(j-i2) \right)
\label{AP2}
\eea

As in the previous case {\bf A}, only 
anomalous dimensions $\gamma^{(1)}_{ii}$ are important  
at $O(\hat a^2)$ in N=4 SUSY.
The anomalous dimensions $\gamma^{(1)}_{ik}$ $(i \neq k)$ 
contribute at $O(\hat a^2)$ level only to normalization factors
$f^{i,NLO}_{q,g,\ps}(j,Q^2_0)$.


\section{ Conclusion }

Above we reviewed
the results for the next-to-leading corrections to the kernel of the BFKL
equation 
and to the anomalous dimensions of 
twist-2 operators in the extended $N=4$ SUSY.
The absence of the coupling
constant renormalization in this model leads presumable to the M\"{o}bius
invariance of the BFKL equation in higher orders of the perturbation theory.
The cancellation of
non-analytic contributions proportional to $\delta _{n}^{0}$ and $\delta
_{n}^{2}$ in $N=4$ SUSY is remarkable (such terms contribute to $\omega$
in the
framework of QCD
as it has been demonstrated in \cite{KoLi,wu}). This property could be
a possible manifestation of the integrability of
the reggeon dynamics in the Maldacena model \cite{Malda} corresponding to
the $N=4$ SUSY in the limit $N_{c}\rightarrow \infty $. Note, that in this
model the 
eigenvalues of the LLA pair kernels in the evolution equations for the matrix
elements of the quasi-partonic operators are proportional to $\psi
(j-1)-\psi (1)$ 
\cite{N=0,Dubna}, which means, that the corresponding Hamiltonian
coincides with the local Hamiltonian for an integrable Heisenberg
spin model.
The residues of these eigenvalues at the points $j=-k$ are obtained 
from the BFKL equation by an analytic continuation of the anomalous 
dimensions to negative integer values of the conformal spin $|n|$. 
Therefore the DGLAP equation is not independent from the BFKL equation in 
$N=4$
SUSY and their integrability properties at $N_c \rightarrow \infty $ are
presumably related.

\vspace{1cm} \hspace{1cm} {\Large {} {\bf Acknowledgments} 
}

The authors are supported in part by the INTAS 00-366 grant. 
A. Kotikov thanks the Alexander von Humboldt Foundation for
its support
at the beginning of this work.
L. Lipatov is thankful to the Hamburg University for
its hospitality during the period of time when this work was completed. He
was supported also by the NATO and CRDF grants.  

We are indebted to J. Bartels, V. Fadin, V. Kim, R. Kirschner, R.
Peschansky, V. Velizhanin and the participants of the PNPI Winter School 
for helpful discussions.





\begin{thebibliography}{99}
\bibitem{BFKL}  L.N. Lipatov, Sov. J. Nucl. Phys. {\bf 23} (1976) 338; \\
E.A. Kuraev, 
L.N. Lipatov and V.S. Fadin, Phys. Lett. {\bf B60} (1975) 50; Sov.
Phys. JETP {\bf 44} (1976) 443; 
{\bf 45} (1977) 199.
%
\bibitem{BL}  Ya.Ya. Balitsky and L.N. Lipatov, Sov. J. Nucl. Phys. {\bf 28}
(1978) 822; 
Sov. Phys. JETP Lett. {\bf 30} (1979) 355.
%
\bibitem{DGLAP}  V.N. Gribov and L.N. Lipatov, Sov. J. Nucl. Phys. {\bf 15}
(1972) 438, 
{\bf 15} (1972) 675;\newline
L.N. Lipatov, Sov. J. Nucl. Phys. {\bf 20} (1975) 94;\newline
G. Altarelli and G. Parisi, Nucl. Phys. {\bf B126} (1977) 298; \newline
Yu.L. Dokshitzer, Sov. Phys. JETP {\bf 46} (1977) 641.

\bibitem{H1}  H1 Collaboration, S. Aid {\em et al}., Nucl. Phys. {\bf B470}
(1996) 3;\newline
ZEUS Collaboration, M. Derrick {\em et al}., Zeit. Phys. {\bf C69} (1996)
607.

\bibitem{corAP}  G. Gurci, W. Furmanski and R. Petronzio, Nucl. Phys. {\bf %
B175} (1980) 27;\newline
W. Furmanski and R. Petronzio, Phys. Lett. {\bf B97} (1980) 437.

\bibitem{LF89}  L.N. Lipatov and V.S. Fadin, Sov. J. Nucl. Phys. {\bf 50}
(1989) 712.

\bibitem{FL}  V.S. Fadin and L.N. Lipatov, Phys. Lett. {\bf B429} (1998)
127.

\bibitem{CaCi}  G. Camici and M. Ciafaloni, Phys. Lett. {\bf B430} (1998)
349.
\bibitem{KoLi}  A.V. Kotikov and L.N. Lipatov, Nucl. Phys. {\bf B582} (2000)
19.
\bibitem{conf}  L.N. Lipatov, Sov. Phys. JETP {\bf 63} (1986) 904.
\bibitem{integr}  L. N. Lipatov, Phys. Lett. {\bf B309} (1993) 394, 
preprint
DFPD/93/TH/70, hep-th/9311037.
\bibitem{BKP}  J. Bartels, Nucl. Phys. {\bf B175} (1980) 365;\newline
J. Kwiecinski and M. Prascalowich, Phys. Lett. {\bf B94} (1980) 413.

\bibitem{LFK}  L.\thinspace N. Lipatov, Sov. Phys. JETP Lett. {\bf 59}
(1994) 596;\newline
L. D. Faddeev and G. P. Korchemsky, Phys. Lett. {\bf B342} (1995) 311.

\bibitem{dual}  L. N. Lipatov, Nucl. Phys. {\bf B548} (1999) 328.
\bibitem{bvl}  L. N. Lipatov, in: {\it Proceedings of the International 
Workshop DIS'99}, Zeuthen, 1999, pp. 207-209;\newline
J. Bartels, G. P. Vacca and L. N. Lipatov, 
Phys. Lett. {\bf B477} (2000) 178.
\bibitem{Lipatov}  L.N. Lipatov, Nucl. Phys. {\bf B452} (1995) 369; 
Physics Reports {\bf 320} (1999) 249.

\bibitem{qp}  A. P. Bukhvostov, G. V. Frolov, E. A. Kuraev and L. N.
Lipatov, Nucl. Phys. {\bf B258} (1985) 601.
\bibitem{N=0}  L. N. Lipatov, Perspectives in Hadronic Physics, in: {\it 
Proc. of the ICTP conf.} (World Scientific, Singapore, 1997).

\bibitem{Dubna}  L. N. Lipatov, in: {\it Proc. of the Int. Workshop on
very high multiplicity physics}, Dubna, 2000, pp.159-176; 
Nucl. Phys. Proc. Suppl. {\bf 99A} (2001) 175.

\bibitem{Lewin}  L.Lewin, Polylogarithms and Associated Functions
(North Holland, Amsterdam, 1981).

\bibitem{Devoto}  A. Devoto and D.W. Duke, Riv. Nuovo Cim. {\bf 7} (1984) 1;\\
%
N. Nilsen, Nova Acta {\bf 90} (1909) 125;\newline
K.S. Kolbig, J.A. Mignaco and E. Remiddi, BIT {\bf 10} (1970) 38; Nuovo Cim.
{\bf A11} (1972) 824.



\bibitem{LS3}  A.I. Davydychev and J.B. Tausk, Phys. Rev. {\bf D53} (1996)
7381;\newline
J. Fleischer, M.Yu. Kalmykov and A.V. Kotikov, Phys. Lett. {\bf B462} (1999)
169;
in: {\it 6th Int. Workshop on Software Engineering, Artificial Intelligence, 
Neural Nets, Genetic Algorithms, Symbolic Algebra, Automatic Calculation 
(AIHENP 99)}, Heraklion, Crete, Greece, 12-16 April, 1999 (hep-ph/9905379);\\
A.I. Davydychev, Phys. Rev. {\bf D61} (2000) 087701;\\
A.I. Davydychev and M.Yu. Kalmykov, Nucl. Phys. 
{\bf B605} (2001) 266.


\bibitem{bfklp}  S.J. Brodsky, V.S. Fadin, V.T. Kim, L.N. Lipatov and G.B.
Pivovarov, JETP. Lett. {\bf 70} (1999) 155; 
in: {\it Proceedings of the PHOTON2001}, Ascona, Switzerland, 2001
(CERN-TH/2001-341, SLAC-PUB-9069, hep-ph/0111390);\\
V.T. Kim, L.N. Lipatov and G.B. Pivovarov, 
in: {\it Proceedings of the VIIIth Blois Workshop at IHEP}, Protvino,Russia,
1999 (IITAP-99-013, hep-ph/9911228);
in: {\it Proceedings of the Symposium on Multiparticle Dynamics (ISMD99)}, 
Providence, Rhode Island, 1999 (IITAP-99-014, hep-ph/9911242).
\bibitem{L3}  L3 Collaboration, 
M. Acciarri et al., Phys. Lett. {\bf B453} (1999) 94.

\bibitem{resum}  B. Andersson, G. Gustavson and J. Samuelson, Nucl. Phys.
{\bf B467} (1996) 443;\newline
B. Andersson, G. Gustavson and H. Kharraziha, Phys. Rev. {\bf D57} (1998)
5543;\newline
G. Salam, JHEP {\bf 9807} (1998) 019;\newline
M. Ciafaloni, D. Colferai and G.P. Salam, JHEP {\bf 9910} (1999) 017; Phys.
Rev. {\bf D60} (1999) 114036;\newline
M. Ciafaloni and D. Colferai, Phys. Lett. {\bf B452} (1999) 372;\newline
R.S. Thorne, Phys. Rev. {\bf D60} (1999) 054031;\newline
G. Altarelli, R.D. Ball and S. Forte, Nucl. Phys. {\bf B575} (2000) 313;
{\bf B599} (2001) 383.


\bibitem{KoLiVe}  A.V. Kotikov, L.N. Lipatov and V.N. Velizhanin, 
work in progress.


\bibitem{CheKaTka}  K.G. Chetyrkin, A.L. Kataev and F.V. Tkachov, Nucl.
Phys. {\bf B174} (1980) 345.
\bibitem{KaKo}
D.I. Kazakov and A.V. Kotikov, Theor. Math. Phys. {\bf 73} (1987) 1264; 
Nucl. Phys. {\bf B307} (1988) 721; 
{\bf B345} (1990) 299(E).

\bibitem{Ko96}  A.V. Kotikov, Theor. Math. Phys. {\bf 78} (1989) 134;
Phys. Lett. {\bf B375} (1996) 240;
in: {\it Proceedings of the XVth Int. Workshop ``High Energy Physics and
Quantum Field Theory''}, Tver, Russia, 2000 (hep-ph/0102177).

\bibitem{FleKoVe} J. Fleischer, A.V. Kotikov and O.L. Veretin, Nucl. Phys. 
{\bf B547} (1999) 343; 
Acta Phys. Polon. {\bf B29} (1998) 2611, hep-ph/9808243.
\bibitem{FleKoVe1} J. Fleischer, A.V. Kotikov and O.L. Veretin, Phys. Lett. 
{\bf B417} (1998) 163.

\bibitem{DEM}  A.V. Kotikov, Phys. Lett. {\bf B254} (1991) 158;
{\bf B259} (1991) 314; 
{\bf B267} (1991) 123;
in: {\it Proceedings of the XVth Int. Workshop ``High Energy Physics and
Quantum Field Theory''}, Tver, Russia, 2000 (hep-ph/0102178).

\bibitem{MerNeer}  R. Merting and W.L. van Neerven, Z. Phys.  
{\bf C70} (1996) 625.

\bibitem{AnalCont}  A.V. Kotikov, Phys. At. Nucl.  {\bf 57} (1994) 133.
%
\bibitem{wu}  V. N. Gribov, L. N. Lipatov and G. V. Frolov, Phys. Lett. 
{\bf B31} (1970) 34; Sov. J. Nucl. Phys. {\bf 12} (1971) 543;\\
H. Cheng and T. T. Wu, Phys. Rev. {\bf D1} (1970) 2775;
{\it Expanding Protons: Scattering at High Energies}
(MIT press, Cambridge, Massachusetts, 1987).
%

\bibitem{Malda}  J. Maldacena, Adv. Theor. Phys. {\bf 2} (1998) 231, Int. J.
Theor. Phys. {\bf 38} (1998) 1113. 


\end{thebibliography}
\end{document}